\documentclass[showpacs,preprintnumbers,prd,nofootinbib]{revtex4-1}

\usepackage{amsmath}
\usepackage{amsfonts}
\usepackage{amssymb}
\usepackage{graphicx}
\usepackage{color}

\usepackage{hyperref}
\usepackage{booktabs}
\usepackage{hyperref}
\usepackage{cleveref}
\usepackage{color}

\begin{document}

\title{A thermodynamic approach to holographic dark energy}

\author{Orlando Luongo}	\email{luongo@na.infn.it}
\address{Department of Mathematics and Applied Mathematics, University of Cape Town, Rondebosch 7701,
Cape Town, South Africa.\\Astrophysics, Cosmology and Gravity Centre (ACGC), University of Cape Town, Rondebosch 7701, Cape Town, South Africa.\\School of Science and Technology, University of Camerino, I-62032, Camerino, Italy.\\Dipartimento di Fisica, Universit\`a di Napoli  ``Federico II'', Via Cinthia, I-80126, Napoli, Italy.\\Istituto Nazionale di Fisica Nucleare (INFN), Sez. di Napoli, Via Cinthia 9, I-80126 Napoli, Italy.}

%%%%%%%%%%%%%%%%%%%%%%%%%%%%%%%%%%%%%%%%%%%%%%%%%%%%%%%%%%%%%%%%%%%%%%%%%%%%%%%%%%%%%%%%%%%%
%%%%%%%%%%%%%%%%%%%%%%%%%%%%%%%%%%%%%%%%%%%%%%%%%%%%%%%%%%%%%%%%%%%%%%%%%%%%%%%%%%%%%%%%%%%%
%%%%%%%%%%%%%%%%%%%%%%%%%%%%%%%%%%%%%%%%%%%%%%%%%%%%%%%%%%%%%%%%%%%%%%%%%%%%%%%%%%%%%%%%%%%%
%%%%%%%%%%%%%%%%%%%%%%%%%%%%%%%%%%%%%%%%%%%%%%%%%%%%%%%%%%%%%%%%%%%%%%%%%%%%%%%%%%%%%%%%%%%%
%%%%%%%%%%%%%%%%%%%%%%%%%%%%%%%%%%%%%%%%%%%%%%%%%%%%%%%%%%%%%%%%%%%%%%%%%%%%%%%%%%%%%%%%%%%%

\begin{abstract}
We propose a method to relate the holographic minimal information density to the de Broglie's wavelength at a given universe temperature $T$. To figure this out, we assume that the thermal length of massive and massless constituents represents the cut-off scale of the holographic principle. To perform our analysis, we suppose two plausible universe volumes, i.e. the adiabatic and the horizon volumes, i.e. $V \propto a^3$ and $V \propto H^{-3}$ respectively, assuming zero spatial curvature. With these choices in mind, we evaluate the thermal lengths for massive and massless particles and we thus find two cosmological models associated to late and early cosmological epochs. We demonstrate that both models depend upon a free term $\beta$ which enters the temperature parametrization in terms of the redshift $z$. For the two treatments, we show evolving dark energy terms which can be compared with the $\omega$CDM quintessence paradigm when the barotropic factor takes the formal values: $\omega_0=-\frac{1}{3}(2+\beta)$ and $\omega_0=-\frac{1}{3}(1+2\beta)$ respectively for late and early eras. From our analyses, we candidate the two models as viable alternatives to dark energy determined from thermodynamics in the field of the holographic principle.
\end{abstract}

%%%%%%%%%%%%%%%%%%%%%%%%%%%%%%%%%%%%%%%%%%%%%%%%%%%%%%%%%%%%%%%%%%%%%%%%%%%%%%%%%%%%%%%%%%%%
%%%%%%%%%%%%%%%%%%%%%%%%%%%%%%%%%%%%%%%%%%%%%%%%%%%%%%%%%%%%%%%%%%%%%%%%%%%%%%%%%%%%%%%%%%%%
%%%%%%%%%%%%%%%%%%%%%%%%%%%%%%%%%%%%%%%%%%%%%%%%%%%%%%%%%%%%%%%%%%%%%%%%%%%%%%%%%%%%%%%%%%%%
%%%%%%%%%%%%%%%%%%%%%%%%%%%%%%%%%%%%%%%%%%%%%%%%%%%%%%%%%%%%%%%%%%%%%%%%%%%%%%%%%%%%%%%%%%%%
%%%%%%%%%%%%%%%%%%%%%%%%%%%%%%%%%%%%%%%%%%%%%%%%%%%%%%%%%%%%%%%%%%%%%%%%%%%%%%%%%%%%%%%%%%%%


\maketitle
\section{Introduction}

Almost two decades ago, the universe acceleration has been observed \cite{uno} and gradually confirmed \cite{due,tre}, showing that our universe seems to be dominated by an exotic anti-gravitational component, dubbed dark energy. Even though the universe speed up is today well-consolidate \cite{quattro}, its physical origin has not been clarified. If the cosmological principle holds, one needs to introduce within Einstein's gravity an additional effective fluid which exhibits a negative pressure, dominating over matter today. This is the philosophy of the concordance paradigm, named the $\Lambda$CDM model. In this approach, one makes use of a vacuum energy cosmological constant\footnote{For the sake of clearness, the cosmological constant approach is physically different from dark energy. In the second case, there is no need to introduce the effects due to quantum vacuum energy into Einstein's equations, employing an evolving barotropic factor.} \cite{cinque,sei,sette,otto}, providing a constant equation of state \cite{nove,dieci} which counterbalances the action of gravity after the transition time\footnote{Which represents the onset of acceleration \cite{undici}, i.e. when the cosmological constant starts dominating over pressureless matter.}. In this picture, the universe energy budget is essentially composed by cold dark matter for about $23\%$, by $4\%$ of baryons as visible matter and by $73\%$ of cosmological constant \cite{dodici,tredici}. The $\Lambda$CDM model shows an excellent guidance with current data \cite{quattordici,quindici,sedici,diciassette,diciotto} but it is severely plagued by the coincidence and fine tuning problems \cite{diciannove,venti}. These issues do not enable to conclude that the $\Lambda$CDM model represents the final paradigm to describe the universe dynamics\footnote{In particular, the measures of cosmological constant and matter magnitudes are extremely close each other, i.e. $\Omega_\Lambda/\Omega_m\sim 2.7$, at our
time \cite{ventuno}, whereas the predicted and observed cosmological constant values differ for about 123 orders of magnitude \cite{ventidue}.}. As a natural consequence, a wide amount of cosmological models have been introduced to extend the concordance model in order to get possible alternatives to dark energy derived from prime principles (see for example \cite{ventitre} and references therein).

Among all, an enticing prospect is to postulate the validity of the holographic principle \cite{ventiquattro}. The principle is supported by the fact that the maximum entropy inside a physical region may be not extensive, growing as the surface area. In this landscape, a cut-off scale is naturally introduced, being responsible for the cosmic acceleration today. In cosmology, one can therefore assume the existence of a minimal information region, consistent with current
cosmological constant's value, but physically different from vacuum energy. Thereby, the understanding of dark energy's nature is shifted to find how the aforementioned cut-off scale evolves in terms as the universe expands. Several different possibilities have been investigated during the last years \cite{ventisette}. In particular, a remarkable class of models is likely represented by choosing cut-off scales which reproduce the $\Lambda$CDM behavior at low and high-redshift regimes \cite{ventotto}. In such a way, the corresponding holographic dark energy naturally reduces to the $\Lambda$CDM paradigm at
late and early times respectively, in agreement with observations and without concordance and fine tuning issues.
We here present how to obtain the holographic cut-off assuming as size of minimal information the universe's thermal
lengths, hereafter $\lambda_{Th}$, associated to massive and massless particles. We propose that the holographic dark energy
behaves like a gas, whose cut-off length corresponds to the standard de Broglie's wavelength at a given temperature
$T$ \cite{ventinove}. For massive and massless particles we have $\lambda_{Th,1}=h(2\pi m k_B T)^{-0.5}$ and $\lambda_{Th,2}=h(2\pi^{\frac{1}{3}} k_B T)^{-1}$ respectively
 \cite{trenta}. In particular, as $\lambda_{Th}$ becomes much smaller than the distances among particles, the dark energy gas obeys the
Maxwell distribution, showing instead quantum effects in the opposite case. This implies the existence of two regimes
in which one can choose, at small scales, quantum effects associated to photon, whereas classical behaviors at large
scales, associated to matter.
The importance of our approach lies on the fact that considering the validity of pure thermodynamics
and that the holographic principle holds at cosmic scales, as two intrinsic elements of Einstein's theory,
it is possible to recover effective dark energy scenarios, instead of postulating the dark energy form
by hand in Einstein's equations and without modifying general relativity.
To guarantee that the holographic cut-off is due to the thermal lengths, we assume the temperature parameterized
by the scale factor as simple as possible, by departing from the standard temperature of the Cosmic Microwave
Background (CMB) in terms of a simple power law. The consequences are here described at both scales. We
investigate how the cosmological models evolve with respect to the redshift $z$ and we fix theoretical limits over the
free parameters of our approach. Moreover, we check the goodness of our holographic dark energy by means of first
order perturbation equations and we compare our treatment with the $\Lambda$CDM and with the quintessence models.
The paper is structured as follows: in Sec. II, we briefly report how to relate and motivate the holographic principle
to cosmology. In Sec. III, we discuss the method to build up the holographic cut-off in terms of the thermal length at
both late and early regimes. In Secs. III A - III B, we discuss some consequences of our choices at the two investigated
stages of universe's evolution. Finally, Sec. IV is devoted to conclusions and perspectives of our work.

\section{The approach of holographic dark energy}

A consolidate conjecture states that general relativity may break down as it is applied to small distances in the
regime of very high energies. The level of small distances and/or high energies may cause the breakdown of geometrical
continuum as well. Examples are offered by lattice or grids in which the interactions are modified either at classical
or quantum levels \cite{trentuno,trentadue}. Assuming quantum effects, in analogy to what happens in quantum mechanics, there
would exist indivisible units with corresponding minimal lengths associated to certain physical domains. If a minimal
length exists, namely $\mathcal L$, one then expects a corresponding minimal area, naively proportional to $\mathcal L^2$. In this picture,
a standard hypothesis is to imagine that thermodynamics does not break down. In so doing, the associated entropy is
modified by the existence of the aforementioned minimal surface and the Bekenstein-Hawking surface law is still valid.
This scheme has been severely supported even by other frameworks which are in favor of the existence of minimal
lengths, among them: the generalized uncertainty principle, string scenarios, doubly special relativity and so forth
\cite{trentatre}.
Another approach which predicts a minimal length is based on the holographic principle. The principle has been
firstly discussed by 't Hooft and extended to other formalisms, such as string theory by Susskind \cite{trentaquattro}. It states that
the degrees of freedom of a spatial region reside not in the bulk but in the boundary. The number of boundary degrees
of freedom per Planck's area should not be larger than one. So that, the basic demand lies on postulating that the
maximum entropy inside a physical region is not extensive and, in particular, that it grows as the surface area.
Exceptional emphasis has been devoted to applying the principle in the context of homogeneous and isotropic
universe \cite{ventiquattro}. In this landscape, the byproduct of extending the holographic postulate to cosmology imposes the
minimal information density as the source of dark energy. Indicating dark energy's density with $\rho_X$, we find that it
is proportional to an infrared cut-off scale, equivalent to $\mathcal L$. From the above perspective, the effective dark energy
becomes:

\begin{equation}\label{holodens}
\rho=\frac{\mathcal A}{\mathcal L^2}\,,
\end{equation}

with $\mathcal A$ a constant which is plausibly close to the present critical density\footnote{Hereafter $8\pi G=1$.}, $\rho_{cr} = 3H^2_0$. This can be viewed as the existence of an information density entering the Einstein's energy momentum tensor. So that, for a precise choice of $\mathcal L$, the universe can accelerate in agreement with recent observations. Thereby, the problem of understanding the
dark energy nature is practically shifted to defining somehow $\mathcal L$, although the physical nature is however completely
reinterpreted with respect to the cosmological constant. To define $\mathcal L$, probably the most appealing and suggestive
frameworks take the size $\mathcal L$ inside the class of functions able to describe unified dark energy models, in which dark
energy emerges as an effect of dark matter. Although reasonable, this procedure generated a wide number of models
which have been built up ad hoc \cite{ventisette,trentacinque,trentasei}.
A simple and consolidate strategy is to assume L proportional to the Ricci scalar or to root mean square of second
order geometrical invariants \cite{ventotto}. This puts in correlation $\mathcal L$ with invariant quantities leading to a geometrical infrared
cut-off scale, satisfying the problem of causality, portrayed in \cite{trentasette}. Dark energy is thus interpreted as a geometrical
acceleration, in strict analogy with the classes of modified gravities which make use of higher order curvature invariants
\cite{trentotto}. These approaches differ from choosing cosmological distances as $\mathcal L$ cut-off scales and candidate to be
invariant. Even though invariant, the above picture does not however take into account the universe thermodynamics
in terms of geometry. Indeed, more recently the idea that thermodynamics may be obtained by means of corresponding
spaces of equilibrium in which the states identify the properties of the system in terms of geometric equilibrium has
been investigated \cite{trentanove}.
Thus, motivated by the idea that there would exist a correspondence between thermodynamics and geometry, one
would expect that the cut-off could be represented by thermal lengths. The thermal length provides a direct measure
of the thermodynamic uncertainty for a thermodynamic system. In other words, supported by the fact that universe
constituents show a thermal scale due to their equilibrium temperatures, it is licit to presume that the cut-off size is
imposed by the temperature itself and by its evolution. The main consequences of treating the cut-off as a thermal
minimal size may give relevant information on dark energy's evolution. In the next section, we describe the two
possible thermal lengths, associated to massive (dust) and non-massive (photons) universe constituents. We show the
simplest case in which dark energy is modeled as perfect gas, in which the holographic cut-off is proportional to the
CMB temperature and we suppose a power law dependence on the redshift $z$, parameterized by an additional free
constant $\beta$.

\section{Thermal cut-off scales and consequences in cosmology}

As above stated, the holographic principle certifies the existence of a further energy density, hereafter $\rho_X$, which
is associated to the minimal information density of a given physical region. In particular, the term $\rho_X$ turns out to
be different from the standard pressureless matter density, $\rho_m$, and may manifest under certain conditions a negative
pressure. We here assume that $\rho_X$ naturally enters the energy momentum tensor of Einstein's gravity, without
adding quantum vacuum energy associated to the cosmological constant. To do so, we employ the validity of the
cosmological principle. So that, the large scale geometric and physical properties can be accounted adopting the
Friedman-Robertson-Walker (FRW) line element:

\begin{equation}\label{metricafrw}
ds^2=dt^2-a(t)^2\Big[dr^2+r^2(\sin^2\theta d\phi^2+d\theta^2)\Big]\,,
\end{equation}

\noindent with zero spatial curvature\footnote{The generalization to $\Omega_k\neq 0$ is straightforward.}, $\Omega_k = 0$. For a source given by a perfect fluid, the whole universe dynamics is expressed in terms of the Friedman equations:

\begin{subequations}\label{equazionifriedmann}
\begin{align}
H^2&=\frac{\rho}{3}\,,\\
\dot H+H^2&=-\frac{1}{6}(\rho+3P_X)\,,
\end{align}
\end{subequations}

\noindent which determine the universe evolution at all stages. The dots represent the derivative with respect to the cosmic
time $t$ and matter is supposed to be pressureless, i.e. $P_m = 0$. In this scheme, only perfect fluids are source of
gravitational field and the total density is essentially composed by $\rho=\rho_m+\rho_X+\ldots$ where at first approximation
one can neglect radiation, neutrinos, relics and so forth. Our main idea is to model the whole universe evolution
as a thermodynamic system \cite{quaranta}. Under this hypothesis, the laws of thermodynamics appear to be mathematically consistent with an isotropic and homogeneous geometric background \cite{quarantuno}. In particular, thermodynamic laws
in which arbitrary functions of time are present in spite of the lack of symmetry are compatible with
equations of state which have to be imposed on the resulting solution.
For instance, it is possible to use the first law of thermodynamics to get expressions for the universe temperature
which may furnish numerical results in agreement with recent CMB observations \cite{quarantadue,quarantatre,quarantaquattro,quarantacinque}. By relating the holographic principle to the Friedmann equations \eqref{equazionifriedmann} and by Eq. \eqref{holodens}, one has:

\begin{equation}\label{omegatotale}
\Omega_{tot}\rho_{cr}=\rho_{m,0}a^{-3}+(1-\rho_{m,0})\mathcal L(a)^{-2}\,,
\end{equation}

\noindent where we guarantee that $H(z = 0) = H_0$ today invoking that $\frac{\mathcal A}{\rho_{cr}}=1 - \rho_{m,0}$. The problem of determining the dark
energy evolution is shifted to reconstructing $\mathcal L(a)$ at different stages in terms of the scale factor, or equivalently by
using the redshift $z$. In other words, since $\mathcal L(a)$ is not known \emph{a priori}, one passes from postulating a dark energy
evolution to understanding which cosmological size can be related to the holographic principle.
The most general holographic dark energy model, i.e. the one in which no specific cut-offs have
been proposed, has been severely investigated and Eq. \eqref{omegatotale} effectively represents a precise class of
models inside the most general ones. The choice of Eq. \eqref{omegatotale}, indeed, enters the set of models in which
there exists a split between matter and holographic dark energy with no interactions between them.
Moreover, the cut-off scale is supposed to be analytical with respect to the redshift $z$. These naive
conditions have been imposed to be consistent with current knowledge on dark energy's evolution and
to guarantee that at the very small redshift regime the holographic models may reduce to the effective
$\Lambda$CDM paradigm. All holographic dark energy models, such as Ricci dark energy models, second order
invariants and so forth are contained in the above classes of generalized holographic frameworks.
In particular, to be consistent with observations, we suppose that dark energy departs smoothly from a pure
cosmological constant. This is mostly supported by observations that suggest that either at late or early times dark
energy slightly departs from a pure constant. In other words, whatever form of dark energy density one considers,
the prescription at late and early times may guarantee that:

\begin{subequations}\label{condiz}
\begin{align}
\rho_{X,0}\approx1-\rho_{m,0}\qquad\qquad at\,\,\,late\,\,\,times\,,\nonumber\\
\,\\
\rho_{X,\infty}\approx1-\rho_{m,0}\qquad\qquad at\,\,\,early\,\,\,times\,,\nonumber
\end{align}
\end{subequations}

For example, in the case of Chevallier-Polarski-Linder parametrization \cite{quarantasei}, the equation of state for dark energy
becomes: $\omega = \omega_0 + \omega_1(1 - a(t))$, which turns out to be a constant at late and early times, giving respectively
$\omega_{late} = \omega_0$ and $\omega= \omega_0 + \omega_1$.
The second imposition may be stringent for models in which dark energy evolves faster than $a^{-3}$ at high redshift.
Hence, one can imagine to overcome this problem, assuming that the reconstructed dark energy term satisfies at least:

\begin{equation}\label{omegaXdiz}
\frac{\Omega_X(z)}{(1+z)^3}\sim\Omega_{m,0}\,,
\end{equation}

with $\Omega_X\equiv\rho_x/\rho_{cr}$, $\forall z$.

Together with the requirements of Eq. \eqref{condiz}, one needs that standard thermodynamics is not violated due to fast
transitions or anomalous evolutions of $\rho_X$ during intermediate phases, i.e. $0 < z < \infty$. This request excludes possible
dark energy oscillation or anomalous behavior and in particular provides that dark energy smoothly evolves, suggesting
that the CMB temperature would be a simple power law in a FRW universe. Moreover, we suppose that dark energy
behaves like a perfect gas, having that it does not interact with any other fluid constituents. All these requirements
are essentially certified by the great goodness in comparing cosmic data coming from the last scattering surface with
respect to the temperature observed today.
Assuming that the universe is composed by massive and massless particles, such as dust and photons respectively,
and considering a local thermal equilibrium, it is natural to expect a thermal length associated to each constituents.
In particular, the thermal length is built up by using the de Broglie's wavelength in a fluid configuration at a given
temperature. For an ideal gas, the constituents are disposed with average distances among them. If such a distance,
namely $\lambda_{space}$, is due to the length determined from the volume occupied by the gas, one thus obtains:

\begin{equation}\label{lambdaspazio}
\lambda_{space}\propto V^{\frac{1}{3}}\,.
\end{equation}

Clearly, we need the correct universe's volume in order to characterize $\lambda_{space}$. We notice that $V = V_0\,a^3$ fulfills
the weak energy condition and represents the simplest choice to model the volume evolution in terms of the scale
factor. Here, one can imagine that the volume evolution has been built up to the first order approximation with $V_0$
proportional to the universe's radius, i.e. $\frac{1}{H_0}$. However, other approaches have recently suggested that the volume
scales by means of the apparent horizon \cite{quarantasette}, leading to a volume $V \propto H^{-3}$. The former approach defines plausible
causal regions with entropy proportional to $H^{-2}$. Analogously, at domains of large or small redshifts one can first
approximate the volume with $V \propto a^3$, without contradicting the causality condition \cite{quarantotto}.
To match thermodynamic properties within the holographic principle, one can suppose that the cut-off scale $\mathcal L$
is defined by taking the thermal length of a given massive or massless gas. As discussed above, this is physically
supported by the thermal length definition itself. In particular, the thermal length is essential in thermodynamic
systems, since it represents a direct measure of the thermodynamic uncertainty, leading to a localization of a given
particle set, of known mass, with the average thermal momentum.
Hence, from the definition of thermal length it follows that the holographic gas behaves classically only if the de-Broglie's wavelength is much smaller than particle distances. For the two cases, the particle distances in the FRW universe are:

\begin{subequations}\label{lspace1}
\begin{align}
\lambda_{space}&\propto a\,,\\
\,\nonumber\\
\lambda_{space}&\propto H^{-1}\,,
\end{align}
\end{subequations}

respectively for the adiabatic volume $V \propto a^3$ and for the horizon volume $V \propto H^{-3}$. So that, using the approximation of perfect gas, one has:

\begin{equation}\label{ltermica1}
\lambda_{Th,1}=\frac{\alpha_1}{\sqrt{T}}\,,
\end{equation}

and

\begin{equation}\label{ltermica2}
\lambda_{Th,1}=\frac{\alpha_2}{T}\,,
\end{equation}

which are valid for massive particles and massless gases respectively, with $T$ the observable temperature of the universe, i.e. the temperature we observe from the CMB experiments \cite{ventinove,trenta}. The two constants $\alpha_{1;2}$ have been explicitly reported in Sec. I.
As confirmed by observations, a suitable landscape is to assume that at late times the fluid characterizing dark
energy is classical, whereas possible quantum effects may be revealed only at early times, with higher energy scales.
At a first glance, it behooves us that at $z \gg 1$ the presence of photons suggests to use Eq. \eqref{ltermica2}, as thermal length,
whereas the opposite case is expected at $z \ll1$, by making use of Eq. \eqref{ltermica1}.

This implies that Eq. \eqref{ltermica1} might be used at small redshift to guarantee a classical regime, while Eq. \eqref{ltermica2} at early
times to enable possible quantum effects. In other words, two cases are in principle possible, at different redshift
regimes, leading to different scenarios. We summarize them as follows\footnote{With the convention $V_0=1$.}:

\begin{equation}\label{diseguaglianza1}
\lambda_{Th,1}\leq \lambda_{space}\,,
\end{equation}

\begin{equation}\label{diseguaglianza2}
\lambda_{Th,2}\geq \lambda_{space}\,,
\end{equation}

Since, as stated, these choices are valid for classical and quantum regimes respectively, we have:

\begin{subequations}\label{diseguaglianza1bis}
\begin{align}
\alpha_1^{-2}T_{z\ll1}&\geq a^{-2}\,,\\
\,\nonumber\\
\alpha_2^{-1}T_{z\gg1}&\leq a^{-1}\,,
\end{align}
\end{subequations}

\noindent in the case of adiabatic volume and:

\begin{subequations}\label{diseguaglianza1bis1}
\begin{align}
\alpha_1^{-2}T_{z\ll1}\geq H^{2}\,,\\
\,\nonumber\\
\alpha_2^{-1}T_{z\gg1}\leq H\,,
\end{align}
\end{subequations}

\noindent in the case of apparent volume. In all the aforementioned frameworks one can use the well-consolidate temperature
parametrization \cite{quarantanove}:

\begin{equation}\label{temperatureparameterization}
\frac{T}{T_0}=a^{\beta-1}\,,
\end{equation}

which fulfills the requirements that we mentioned above. In principle, modifying the temperature definition with the
parametrization \eqref{temperatureparameterization} would modify the use of the cosmic distances, increasing the incidence of the duality problem
\cite{cinquanta} between the luminosity and angular distances. The request that the duality problem does not significatively
modify observations can be accounted by assuming that $\beta$ is close to $\beta \sim 1$ and may be parameterized by

\begin{equation}\label{betalimit}
\beta=1+\delta\,,
\end{equation}

with $\delta \geq 0$ defined as a small positive number. With the above recipe, we can now consider some consequences of our
approaches to late and early times in the next subsections.

\subsection{Late-time evolution}

The two thermal lengths lead to approaches at different stages of the universe evolution. The first analysis can be
performed by using the adiabatic volume at the small redshift regime. This epoch corresponds to our time, in which
making use of the thermal length for massive particles, one finds:

\begin{equation}\label{Eadiabatico}
\mathcal E^{adiabatic}_{late}=\Omega_{m,0}(1+z)^3+\left(1-\Omega_{m,0}\right)(1+z)^{1-\beta}\,,
\end{equation}

where $\mathcal E \equiv \frac{H}{H_0}$ is the normalized Hubble rate. Using the temperature parametrization of Eq. \eqref{temperatureparameterization} and requiring $z\simeq0$,
one finds

\begin{equation}\label{alfa1maggiore}
\alpha_1\geq \sqrt{T_0}\,,
\end{equation}

which certifies that $\alpha_1$ gives rise to the lowest limit corresponding to the value $\alpha_1\simeq1.65$K. Combining Eq. (1) with Eq. \eqref{temperatureparameterization}, assuming $\alpha_1 = \sqrt{T_0}$, we have:

\begin{equation}\label{OmegaXcostretta}
\Omega_{X,0}\equiv1-\Omega_{m,0}=\frac{\mathcal A}{\rho_{cr}\alpha_1^2}T_0\,,
\end{equation}

and using $\Omega_{m,0}=0.27$ and $\alpha_1\sim T_0$, we simply get $\mathcal A > 0$, as requested by modeling the dark energy density in Eq. \eqref{holodens}, and moreover:

\begin{equation}\label{parametroA}
\mathcal A=\left(1-\Omega_{m,0}\right)\rho_{cr}\,,
\end{equation}

which confirms that $\mathcal A$ is close to the critical density.
The cosmological model depends upon two parameters, i.e. $\Omega_{m,0}$ and $\beta$, and mimics the $\Lambda$CDM behavior at the
level of background cosmology, providing a dark energy term which can be approximated at first order as:

\begin{equation}\label{primordineDE}
\Omega_X\simeq \left(1-\Omega_{m,0}\right)\Big[1+(1-\beta)z\Big]\,,
\end{equation}

reproducing at $z = 0$ the exact value of $\Omega_\Lambda=1-\Omega_{m,0}$, in perfect agreement with observations. In particular, the first
correction to the cosmological constant case, i.e. $(1-\Omega_{m,0})(1-\beta)z$, turns out to be small if compared to $1-\Omega_{m,0}$,
when $\beta<1$.
At higher redshifts, the model mimics the $\omega$CDM paradigm, which corresponds to a quintessence field, whose dark
energy contribute scales as $(1+z)^{3(1+\omega)}$. In this picture, one finds that the two models are formally equivalent if one
imposes:

\begin{equation}\label{eosfinale}
\omega=-\frac{1}{3}(2+\beta)\,.
\end{equation}

The $\omega$CDM model is thought as the simplest approach, generalizing the concordance paradigm, but it is derived
from postulating a quintessence term, built up in terms of a slowly-rolling dynamical scalar field. In this scheme, one
needs to invoke a non-canonical kinetic term in the Lagrangian. Our treatment predicts the same dynamics, without
invoking a quintessence scalar field, i.e. without introducing by hand another ingredient within Einstein's equations.
Hence, although the model here presented can be formally compared with the $\omega$CDM approach,
the physics associated to it turns out to be completely different. The need of a different origin of
quintessence, indeed, enables one to avoid the introduction of scalar fields in the Einstein's energy
momentum tensor. The need of a precise cut-off scale is essentially enough to extend the concordance
paradigm by means of a running dark energy term in which the barotropic factor is constant. From
those considerations, one does not need to expect that the propagation of perturbations is equivalent
to the $\omega$CDM approach, as we discuss later in the work. On the contrary, the pressure, energy density
and equation of state of our model may be formulated in terms of $\beta$ only, which effectively works as a thermodynamical scale. The principal thermodynamic quantities of our model are summarized
as:

\begin{subequations}\label{letreequazioni}
\begin{align}
\rho_X&=\rho_{cr}\left(1-\Omega_{m,0}\right)a^{-1+\beta}\,,\\
P_X&=-\frac{\rho_{cr}}{3}\left(1-\Omega_{m,0}\right)(2+\beta)a^{-1+\beta}\,,\\
S&\propto \frac{\rho_X\mathcal V}{T}=V_0\rho_{cr}\left(1-\Omega_{m,0}\right)a^3\,,
\end{align}
\end{subequations}

\noindent where the last relation suggests that the entropy, $S$, scales as the volume, giving rise to the important
fact that the entropy density is constant, as one expects in the concordance model, but without invoking
a cosmological constant \emph{a priori}.
Moreover, since observations show that $\omega \simeq -1$, it turns out that $\beta\sim1$ and $\delta \sim 0$, confirming what we previously
discussed in Eq. (16).
The same procedure can be performed in the case of the apparent volume, leading to the same Hubble evolution of
Eq. (17). However, in this case one finds:

\begin{equation}\label{alfx}
\alpha_1\leq\frac{T_0}{H_0^2}\,,
\end{equation}

which corresponds to an upper limit for $\alpha_1$ which is positive in order to guarantee that $\lambda_{Th}>0$. Particularly, for $\alpha_1 = \frac{T_0}{H_0}$, one gets:

\begin{equation}\label{Ab}
\mathcal A=H_0^{-1}\left(1-\Omega_{m,0}\right)\rho_{cr}=3\left(1-\Omega_{m,0}\right)\,.
\end{equation}

In both cases, (20) and (27), $\mathcal A$ is of the same order of $1-\Omega_{m,0}$, but in the second case is much more close to $\rho_{cr}$, as
requested by the holographic principle when building up Eq. (1). Thus, the universe accelerates when the quantity $\frac{\dot H}{H^2}$ determines a deceleration parameter's evolution inside the interval $-1 \leq q_0 \leq 0$. For the model $\mathcal E_{late}^{adiabatic}$, for both the
adiabatic and horizon volumes, using $a \equiv (1 + z)^{-1}$ and

\begin{equation}\label{cosmichrono}
\frac{dz}{dt}=-H(z)(1+z)\,,
\end{equation}

\begin{figure}
\begin{center}
\includegraphics[width=3.2in]{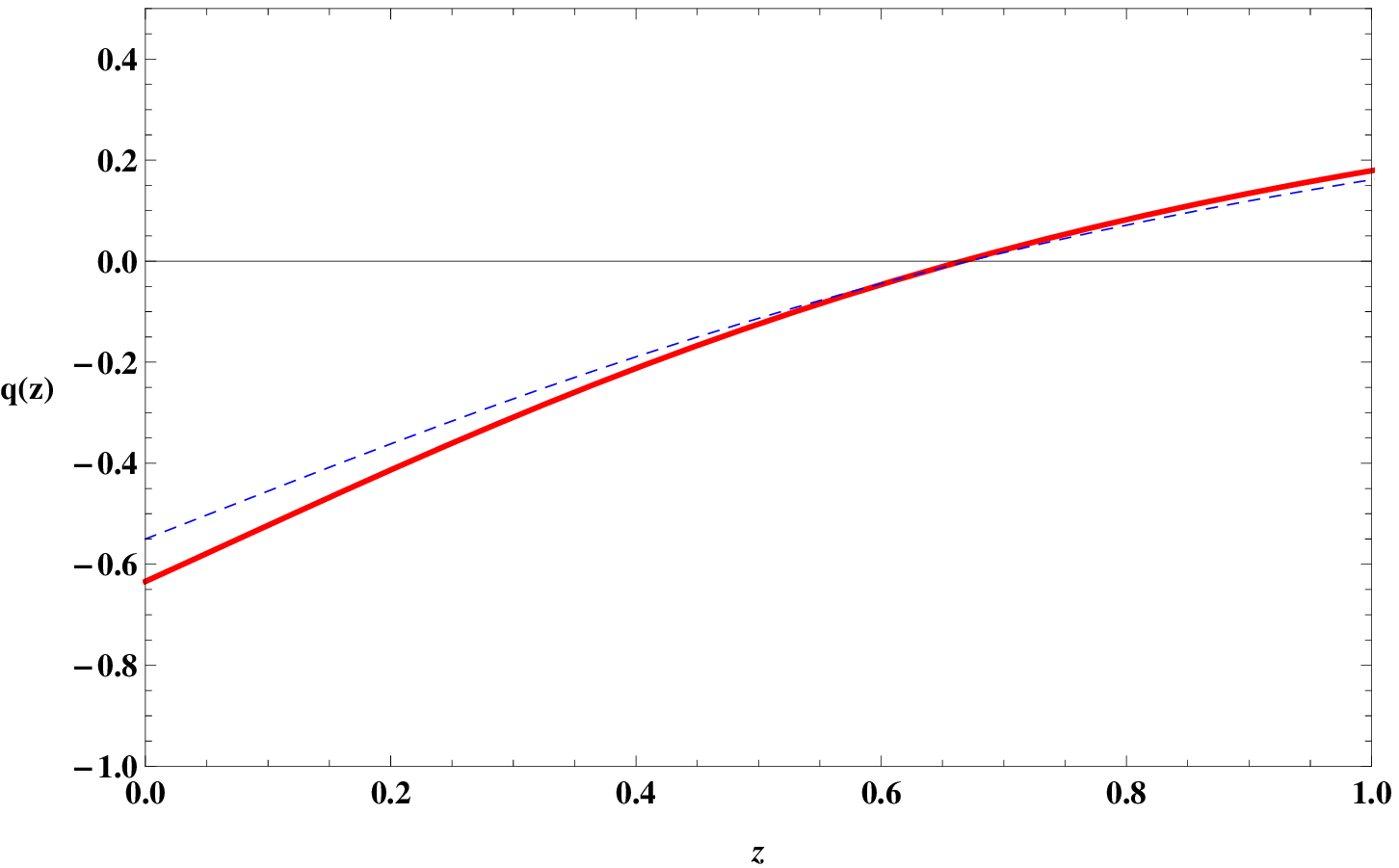}
\includegraphics[width=3.2in]{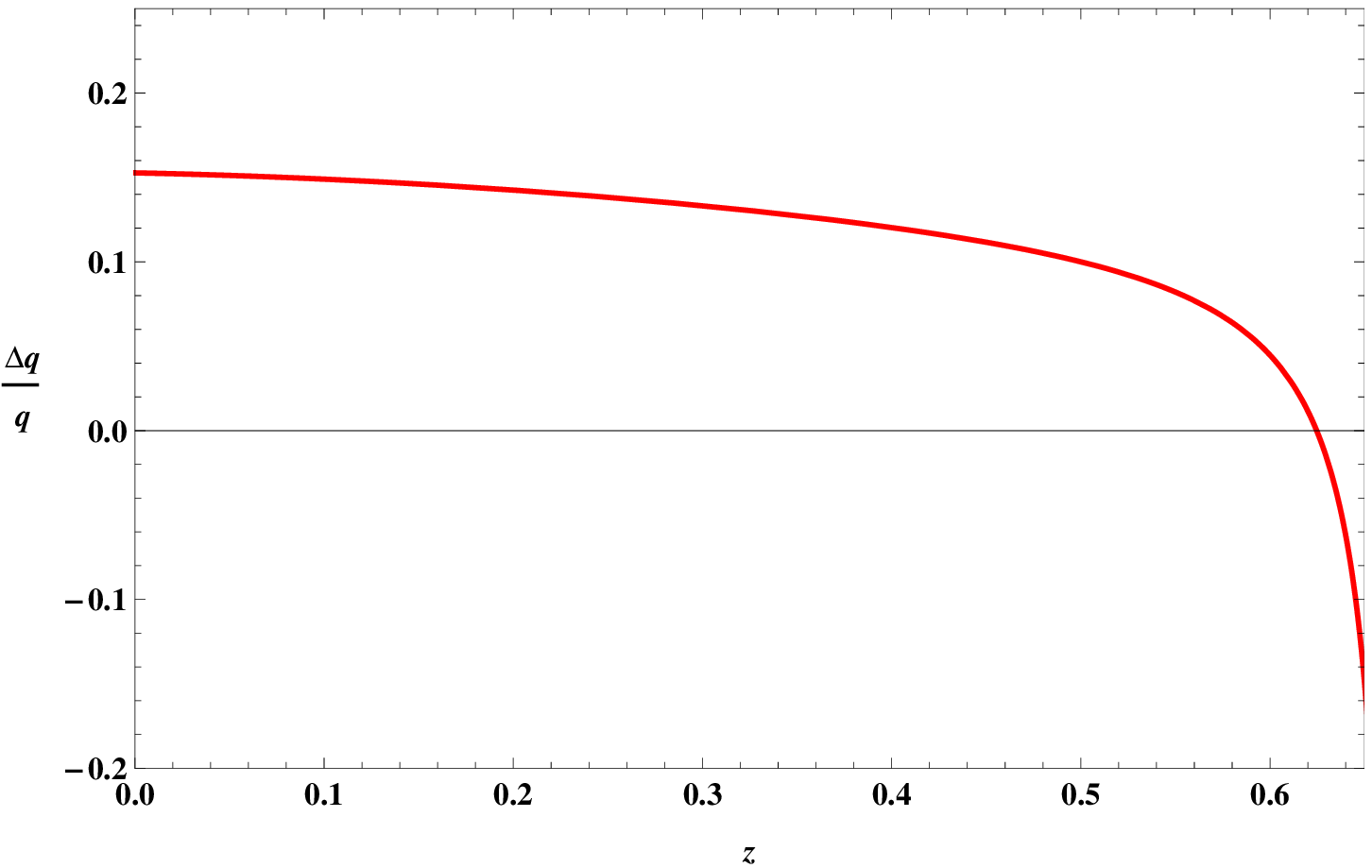}
\caption{Left plot: Behaviors of the deceleration parameters for our model (red line) and for the concordance paradigm (dashed line), with indicative values $\Omega_{m,0}= 0,3$ and $\beta$ as reported in Eq. \eqref{betanumerico}. Right plot: variations between our model and the $\Lambda$CDM approach.}
\label{fig:f_z}
\end{center}
\end{figure}

\noindent we get:

\begin{equation}\label{qvelato}
\frac{\dot H}{H^2}=\frac{3\Omega_{m,0}(1+z)^{2+\beta}+\left(1-\Omega_{m,0}\right)(1-\beta)}{2\Big[1+\Omega_{m,0}(-1+(1+z)^{2+\beta})\Big]}\,,
\end{equation}

\noindent which enters the present value of the deceleration parameter as:

\begin{equation}\label{qoggi}
q_0=\Omega_{m,0}-\frac{1}{2}\Big[1+\frac{\beta}{2}\left(1-\Omega_{m,0}\right)\Big]\,.
\end{equation}

Inside the observational domains, predicted by model independent measurements of the deceleration parameter, i.e. $q_0 = 0.6361^{+0.3720}_{-0.3645}$ \cite{cinquantuno}, we obtain:

\begin{equation}\label{betanumerico}
\beta=1.2460^{+1.2947}_{-1.2733}\,,
\end{equation}

where we considered for the matter density an indicative value: $\Omega_{m,0}=0.30^{+0.05}_{-0.05}$, with errors in Eq. \eqref{betanumerico} propagating
by the standard logarithmic formula, as follows:
\begin{figure}
\begin{center}
\includegraphics[width=3.2in]{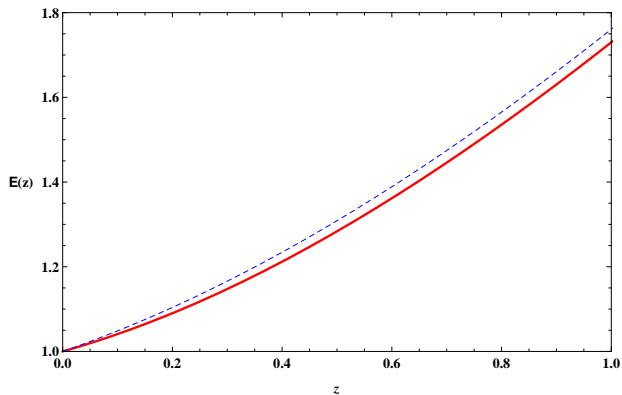}
\caption{Behaviors of our model and $\Lambda$CDM normalized Hubble rates $\mathcal E$, with indicative values $\Omega_{m,0}= 0.3$ and $\beta$ as reported in Eq. \eqref{betanumerico}.}
\label{fig:f_z}
\end{center}
\end{figure}

\begin{equation}\label{deltabeta}
\delta \beta = |\partial_{q_0}\beta|\delta q_0+
|\partial_{\Omega_{m,0}}\beta|\delta \Omega_{m,0}\,,
\end{equation}

where $\partial_{q_0}$ and $\partial_{\Omega_{m,0}}$ indicate the partial derivatives with respect to $q_0$ and $\Omega_{m,0}$ respectively.
Those results are compatible with the $\Lambda$CDM paradigm \cite{cinquantadue,cinquantatre} and correspond to a barotropic factor inside the
interval:

\begin{equation}\label{omeganums}
\omega=-1.0820^{+0.4316}_{-0.4244}\,,
\end{equation}

which testifies the goodness of our approach, lying on the expected intervals capable of accelerating the universe
today.
For the sake of completeness, in Figs. 1, we plot the behaviors of the deceleration parameter of our model versus
the deceleration parameter in the $\Lambda$CDM framework. They evolve similarly in the regime of small redshifts, as one
can notice from the left figure. In the right plot, we report the percentage of variation, which has been defined as:

\begin{equation}\label{deltaqratio}
\frac{\Delta q}{q}\equiv\frac{q-q_{\Lambda CDM}}{q_{\Lambda CDM}}\,.
\end{equation}

It indicates discrepancies smaller than $20\%$ between the two models, in the plotted redshift domain. This is confirmed
by Fig. 2, in which we compare the behaviors of our model versus the concordance paradigm. In particular, assuming $\Omega_{m,0} = \frac{1}{3}$, for three redshifts appropriately chosen within the interval $z \leq 2$, one has:

\begin{equation}\label{deltaq1}
\frac{\Delta q}{q}=-\frac{2}{3}(1-\beta)\,,
\end{equation}

\begin{equation}\label{deltaq2}
\frac{\Delta q}{q}=-\frac{3}{2}-\frac{5(2+\beta)}{2+2^{2+\beta}}\,,
\end{equation}

\begin{equation}\label{deltaq3}
\frac{\Delta q}{q}=\frac{3}{2}\left(3-\frac{29(2+\beta)}{2+3^{2+\beta}}\right)\,,
\end{equation}

evaluated at $z = 0$, $z = 1$, $z = 2$. The percentage function $\frac{\Delta q}{q}$
increases as $\beta$ increases. In particular, for the indicative
value $\beta = 1.2490$, it is easy to get that ${\Delta q\over q}=
\{16.6\%; 8.83\% ; 4.24\%\}$, for the above set of redshift. At fixed $\beta$, the two
deceleration parameters are much more compatible as the redshift increases.
However, at our time, corresponding to
$z = 0$, the deviation increases significantly if one passes from
$\beta = 1.2490$ to $\beta = 1.5$. This indicates that the favorite
values are inside $\delta\simeq0$ with $\beta > 1$.

\subsection{Early time cosmology}

Analogous considerations may be carried out by investigating the consequences of our recipe at higher redshift
domains. So that, at early times, following Eqs. \eqref{diseguaglianza1} and \eqref{diseguaglianza1bis} we require that $\lambda_{Th}>\lambda_{space}$. From this, it follows that
the normalized Hubble rate is:

\begin{equation}\label{Eadiabatico2}
\mathcal E^{adiabatic}_{early}=\Omega_{m,0}(1+z)^3+\left(1-\Omega_{m,0}\right)(1+z)^{2(1-\beta)}\,.
\end{equation}

In this case, using again the temperature parametrization but at the regime in which $z\gg1$, one obtains again Eq.
(20), which confirms that $\mathcal A$ is close to the critical density. For the horizon volume, the two regimes provide the
same result of Eq. (27).
To be compatible with the concordance paradigm, but differently from the $\omega$CDM case, one needs that the sound
speed:

\begin{equation}\label{cs}
c_s^2=\frac{2\left(1-\Omega_{m,0}\right)(\beta-1)(1+2\beta)}{9\Omega_{m,0}(1+z)^{1+2\beta}-6\left(1-\Omega_{m,0}\right)(\beta-1)}\,,
\end{equation}

\noindent is negligibly small. Imposing the former property on $c_s$, one gets that the sound speed  tends to zero when $z \rightarrow \infty$ as $\beta > -1$, which is supported by
theoretical considerations, since $\beta> 0$. Analogously, $c_s$ cannot exceed the stiff matter limit, i.e. $c_s = 1$. To figure this
out, we summarize the following cases:

\begin{eqnarray}
\left\{
  \begin{array}{ll}
    \beta=1, & \hbox{$c_s=0$;}\\
    0<\beta<1, & \hbox{$c_s\rightarrow0$,} \\
    \beta>1, & \hbox{$c_s\rightarrow0$,}
  \end{array}
\right.
\end{eqnarray}

which are associated to different intervals of $\beta > 0$. Notice that the case $\beta > 1$ implies a sound speed which goes
faster to zero than $0 < \beta < 1$.
Moreover, in analogy to the late time case, one can write down the thermodynamic quantities of
particular interest for such a model, given by:
\begin{subequations}\label{lista1}
\begin{align}
\rho_X&=\rho_{cr}\left(1-\Omega_{m,0}\right)a^{2(\beta-1)}\,,\\
P_X&=-\frac{\rho_{cr}}{3}\left(1-\Omega_{m,0}\right)(2+\beta)a^{2(\beta-1)}\,,\\
S&\propto \frac{\rho_XV}{T}=V_0T_0^{-1}\rho_{cr}\left(1-\Omega_{m,0}\right)H^{-3}\,.
\end{align}
\end{subequations}

With those considerations in mind, at a first approximation one can write the first order perturbation equation
approximating the sound speed with $c_s = 0$. This gives:

\begin{equation}\label{pert}
\ddot \delta+2H\dot\delta-4\pi G\rho_m\delta=0\,.
\end{equation}

When the redshift increases, the matter density influences the whole dynamics and so one can suppose to check
how much the growth factor deviates from the concordance model. To do so, we can define the deviation parameter
$\Delta$ as:

\begin{equation}\label{Ddelltta}
\Delta=1-D(a)D^{-1}_{\Lambda}(a)\,,
\end{equation}

implying that the deviations of our model are smaller than $5\%$, with $D(a) = \frac{\delta}{a}$
 for our model and $D_\Lambda(a) = \frac{\delta_{\Lambda CDM}}{a}$
a for the
concordance model.
We can therefore define in perturbation theory the following growth index definition:

\begin{equation}\label{effedef}
f=\frac{d\ln\delta}{d\ln a}\,,
\end{equation}

whose evolution can be obtain by rewriting Eq. (44), with boundary conditions made by using the last scattering
redshift $z_{LSS} \approx 1089$ and $f(a_{LSS}) = 1$. In particular, we have the growth index equation as:

\begin{equation}\label{groweq}
f^\prime+\frac{f^2}{a}+\Big[\frac{2}{a}+\frac{1}{\mathcal E}\frac{d\mathcal E}{da}\Big]f
-\frac{3\Omega_m}{2\mathcal E^2a^4}=0\,.
\end{equation}

\begin{figure}
\begin{center}
\includegraphics[width=3.2in]{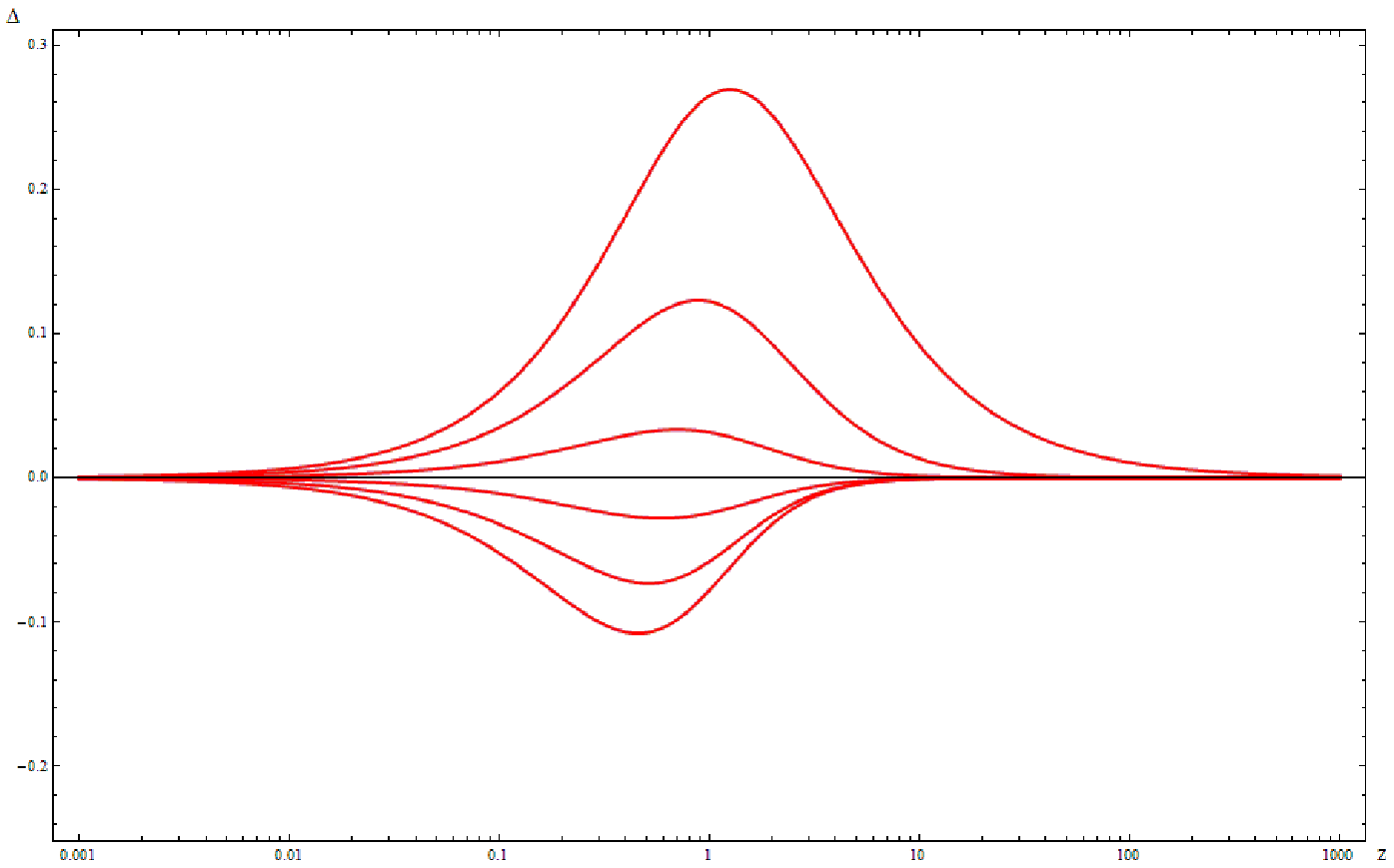}
\includegraphics[width=3.2in]{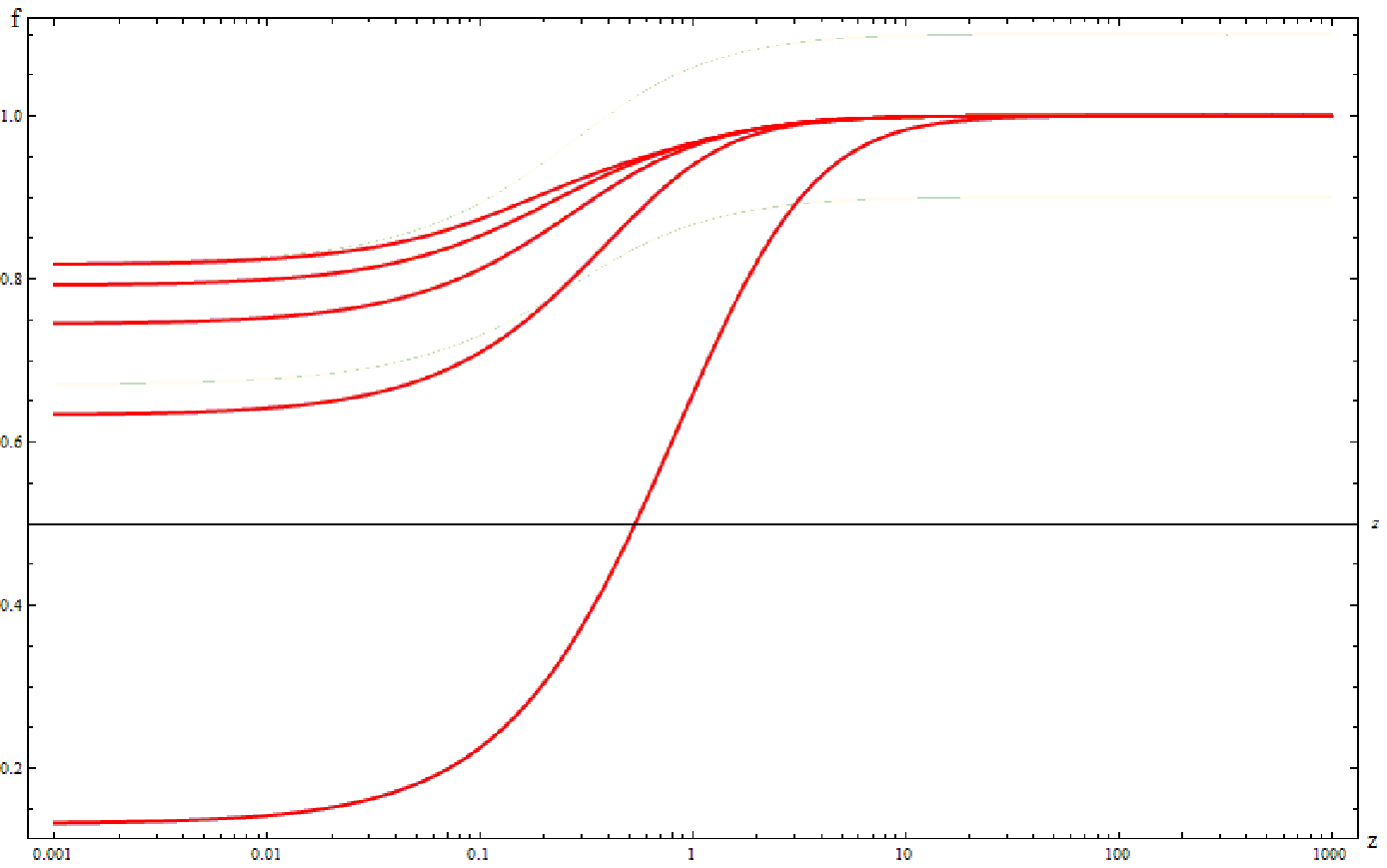}
\caption{In the left plot, we obtain our model with an equidistant grid spacing, split by five steps with the same distance. The
left figure shows slight departures from our framework and the concordance model. This is even confirmed in the second plot,
on the right. There, we report the behavior of the growth factor $f(a)$ in terms of the redshift z. Both redshift domains span
from $z = 0$ to $z \simeq 1000$.}
\label{fig:f_z}
\end{center}
\end{figure}
In Figs. 3, we report the behaviors of $\Delta$ and $f$ for our model. The discrepancies as reported in the above plots are
again non-significative, showing a great compatibility between the standard cosmological model and our paradigm.
A further remark is based on noticing that our approach candidates to predict dark energy at late
stages and even a dark term at earlier epochs of universe evolution\footnote{The model, indeed, does not provide a pure pressureless contribution at higher redshifts and so one cannot conclude that dark energy
is negligibly small when the universe size decreases.}. Naively, any early phase of dark
energy would imply the relation:

\begin{equation}\label{giustoper}
\frac{d}{dt}\left(\frac{1}{aH}\right)<0\,,
\end{equation}

which guarantees the existence of inflation at $z\rightarrow\infty$. Thus, we may argue whether our approach
satisfies the above requirement, giving rise to $\ddot a > 0$ at small scales. From the corresponding equation of state, one immediately concludes that our model works fairly well in predicting a phase in which $\omega<-\frac{1}{3}$, which turns out to represent the minimal condition to guarantee that inflationary time exists
in the past. In other words, the above relation is satisfied and the accelerated behavior of our model
manifests either at late and early phase, providing under certain conditions, that inflation may be
predicted as a byproduct of our thermal lengths, unifying the dark energy description at late times
with the corresponding inflationary epoch at early times.

\section{Final outlooks and perspectives}

In this paper, we assumed the validity of the holographic principle to characterize the dark energy evolution. To do
so, we supposed that the cut-off scale, entering the minimal information density, is given by the well-consolidate de Broglie's
wavelength associated to massive and massless particles, i.e. matter and photons respectively. The corresponding
cut-off scales become the thermal lengths which can be used at late and early times, respectively for classical and
quantum regimes. We built up two cosmological models parameterizing the temperature with respect to the redshift
$z$, using a standard power law, depending upon a constant $\beta$ which enters our approach as a free parameter. To do
so, we chose the adiabatic and the horizon volumes, i.e. $V \propto a^3$ and $V \propto H^{-3}$ respectively. We found that in both
regimes of small and high redshifts, our models mimic the $\Lambda$CDM behavior by means of an evolving dark energy term.
The former dark energy evolution seems to be compatible with the $\omega$CDM quintessence model if the barotropic factor
reads: $\omega_0 = -\frac{1}{3}(2+\beta)$ and $\omega_0 = -\frac{1}{3}(1+2\beta)$ at late and early times. We found a fairly good agreement of our models
with the concordance paradigm. In particular, we found that the deceleration parameter evolution at small redshift
spans within the observational interval predicted by observations and agrees with the $\Lambda$CDM predictions. Moreover,
we found that the shift between our framework and the concordance model turns out to be small at high redshift.
Further, even the growth index evolution slightly departs from the predictions of the standard cosmological model.
This candidates the approach of thermal length within the holographic principle as a viable alternative to dark energy.
In future works, we will better understand whether more complicated models of holographic dark energy may be
obtained, considering thermal length parameterized by specific temperature evolutions. Moreover, we will clarify the
correct ranges of available $\beta$ employing the use of cosmic data, to better understand if $\beta > 1$ or $\beta < 1$. Further, we
will check the goodness of our model at early times with respect to more complete small perturbation analyses.

\end{document}